\documentclass[conference]{IEEEtran}
\IEEEoverridecommandlockouts
\usepackage{cite}
\usepackage{amsmath,amssymb,amsfonts}
\usepackage{algorithmic}
\usepackage{graphicx}
\usepackage{textcomp}
\usepackage{comment}
\usepackage{xcolor}
\def\BibTeX{{\rm B\kern-.05em{\sc i\kern-.025em b}\kern-.08em
    T\kern-.1667em\lower.7ex\hbox{E}\kern-.125emX}}
\newcommand{\linebreakand}{%
  \end{@IEEEauthorhalign}
  \hfill\mbox{}\par
  \mbox{}\hfill\begin{@IEEEauthorhalign}
  }
    
\begin{document}

\title{The Perceptions of Software Engineers Concerning the Utilization of Bots in the OSS Development Process: An Exploratory Survey\\}

\author{\IEEEauthorblockN{Danyellias Vaz de Lima Manso}
\IEEEauthorblockA{\textit{Informatics Institute} \\
\textit{Federal University of Goias, Brazil}\\
Goiania, Goiás \\
danyellias@discente.ufg.br}
\and
\IEEEauthorblockN{Valdemar Vicente Graciano Neto}
\IEEEauthorblockA{\textit{Informatics Institute} \\
\textit{Federal University of Goiás, Brazil}\\
Goiânia, Goiás \\
valdemarneto@ufg.br}
\and
\IEEEauthorblockN{Mohamad Kassab}
\IEEEauthorblockA{\textit{Boston University (USA)} \\
\textit{New York University}\\
Abu Dhabi (UAE) \\
mkassab@bu.edu}
}
 
\maketitle

\begin{abstract}
Software bots, extensively adopted by Open Source Software (OSS) projects, support developers across several activities, from automating predefined tasks to generating code that aids software engineers. However, with the growing prominence of bots, questions have emerged regarding the extension to which they truly assist or hinder software engineers in their routine tasks. To address this, an exploratory survey was conducted with 37 software engineers to gather insights into their views on the use of bots within the software development process. The findings suggest that bots are present across multiple phases of the software development lifecycle, providing daily support to professionals by enhancing productivity and facilitating task automation. Respondents stated that current bots are not sufficiently intelligent and raised new challenges and enhancements to aid bot designers in developing additional functionalities and integrations.
\end{abstract}

\begin{IEEEkeywords}
bot, perception, software engineering, survey, improvements
\end{IEEEkeywords}

\section{Introduction}
Open Source Software (OSS) has fundamentally changed how we understand the Software Development Process \cite{b1}. A successful example is Linux, which demonstrates that collaborative development can produce software of the highest quality, often considered by many developers and users to be superior to paid software. OSS projects are typically developed in a distributed and decentralized way \cite{b2, b3}. 
With the development of large-scale software, it has become increasingly necessary to establish process standards. To maintain such standards and ensure software quality, time-consuming, tedious tasks, or even simple tasks that must be repeated multiple times, are required. These tasks ultimately reduce the productivity of the developers involved in the project. One way to help developers be more productive and effective is to provide them with better and smarter tools, i.e, tools that automate or streamline the development process \cite{b18}. In general, bots are
seen as applications that automate repetitive or predefined tasks \cite{b18}. 

In Software Engineering, bots support several activities, such as communication and decision making \cite{b18}. In collaborative software development environments, bots automate tasks that generally require human interaction \cite{b9}. For example, Urli et al. \cite{b23} propose a bot that serves as a program repair tool-chain for continuous integration build failures in projects hosted on GitHub. Beschastnikh et al. \cite{b24} proposed the use of bots as a solution for the automated deployment and evaluation of software engineering analysis techniques. Wessel et al. \cite{b11} analyzed 351 OSS and found that 93 (26\%) use bots which complement other developers’ work. The authors interviewed project maintainers to investigate, among other things, how contributors and integrators perceive bots during the pull request submission process. 
However, although these studies provide recommendations on how to develop bots, evaluate bot capabilities and performance, as well as the impact of bot usage on pull requests, they do not draw attention to how engineers perceive the support of all types of bots throughout the entire software development process and how this perception may assist in improving the bots. Although there are some studies focused on detecting and understanding the behavior of bots in repositories \cite{b5, b14, b15, b16}, little is known about software engineers' perceptions regarding the use of bots in the software development process.

To gain this insight, a survey was conducted among software engineering professionals through an online questionnaire. Based on the study by Lebtag et al. \cite{b19, b22}, figure 1 illustrates the steps followed in the development of this work, which a questionnaire was planned, the survey was executed with software engineering professionals, the results were analyzed, and a study was elaborated to report the findings.

\begin{figure}[htbp]
\centerline{\includegraphics[scale=0.4]{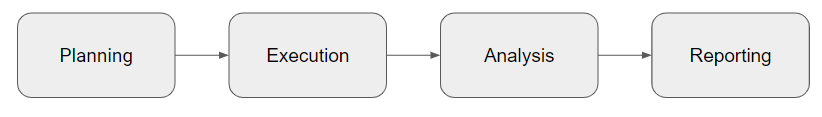}}
\caption{Steps taken for the development of the work.}
\label{fig1}
\end{figure}

The main contribution of this paper is presenting results of an exploratory survey to answer the following research questions: "\textbf{What are the perceptions of software engineers regarding the use of bots in the software development process?}" \textbf{(RQ1)} and "\textbf{How could a bot be evolved to improve the software development process?}" \textbf{(RQ2)}. Thirty-seven software engineers answered the survey. The results obtained indicated that most participating software engineers frequently use bots in the software development process, and a significant portion agree that bots increase productivity within this process. However, several aspects were identified that need improvement in bot development to enable them to adapt better to the context in which they are used and to the user. The remainder of this article is organized as follows: Section 2 describes the research method; Section 3 outlines the conduction and results obtained; Section 4 discusses the limitations of the study, and finally, Section 5 concludes the paper with final remarks.

\section{Research Method}
Survey is a methodology used to obtain knowledge through respondents' opinions and to analyze their responses, attitudes, and behaviors, thereby enabling a better understanding of a specific phenomenon \cite{b20}. Survey research is employed in various fields of knowledge, and in software engineering, it stands out as one of the most widely used methods for conducting exploratory empirical investigations \cite{b21}. The \textbf{goal} of this survey was \textit{to explore software engineers' perspectives on the use of bots in the software development process}. 

For this study, a protocol was developed inspired by the guidelines proposed by Molléri et al. [21] and based on the research flow conducted by Lebtag et al. [19], structured into four steps: Step 1) Planning, Step 2) Execution, Step 3) Analysis, Step 4) Results.

\subsection{Planning}\label{AA}
In step 1, the two research questions (RQ) were established: "RQ1: What are the perceptions of software engineers regarding the use of bots in the software development process?" and "RQ2: How could a bot be evolved to improve the software development process?" To answer these questions, a survey was conducted, and through the analysis of the results, it was possible to address them.

\begin{itemize}
\item \textbf{Identification of Target Audience:} This study’s population is formed by professionals who work on software engineering in either industry or academia.
\item \textbf{Elaboration of the questionnaire:} The objective was to collect perceptions of software engineering professionals on the utilization of bots in the development process via an online questionnaire to enable a broader reach among software engineers.
\end{itemize}

The structure of the questionnaire aligns with Table 1. It consists of (i) demographic questions (DQ), which serve to obtain a profile of the participants, and (ii) opinion questions (OQ), designed to capture the perceptions of software engineering professionals.

\begin{table}[htbp]
\caption{Questionnaire Layout}
\begin{center}
\begin{tabular}{|c|c|}
\hline
\cline{2-2} 
\textbf{Demographic Questions} & \textit{Characterization Questions} \\
\hline
\textbf{Opinions Questions} & \textit{Open-ended and closed-ended questions} \\
\hline
\end{tabular}
\label{tab1}
\end{center}
\end{table}

Closed-ended (CEQ) and open-ended (OEQ) questions were developed. CEQ questions were used to standardize the responses, facilitate reading, and synthesize the necessary information through analytical methods. OEQ allowed for a better description of the participants' understanding and proposals. Tables 2 and 3 present the demographic and opinion questions, respectively.

\begin{table}[htbp]
\caption{Demographic Questions}
\begin{center}
\begin{tabular}{|c|c|p{6cm}|}
\hline
\textbf{ID} & \textbf{Type} & \textbf{Questions}\\
\hline
DQ1 & CEQ & What is your age range?  \\
\hline
DQ2 & CEQ & What country do you reside in?  \\
\hline
DQ3 & CEQ & If you reside in Brazil, which state do you live in?  \\
\hline
DQ4 & CEQ & What is your level of experience in software development? \\
\hline
DQ5 & CEQ & In which field do you currently work? \\
\hline
DQ6 & CEQ & What is the size of the company you work for? \\
\hline
\end{tabular}
\label{tab2}
\end{center}
\end{table}

\begin{table}[htbp]
\caption{Opinion Questions}
\begin{center}
\begin{tabular}{|c|c|p{6cm}|}
\hline
\textbf{ID} & \textbf{Type} & \textbf{Questions}\\
\hline
OQ1 & OEQ & How would you describe what a "bot" is in the context of software development?  \\
\hline
OQ2 & CEQ & How often do you utilize bots in your software development projects?  \\
\hline
OQ3 & CEQ & Assess the following statement according to the options below: "The use of bots enhances productivity in software development."  \\
\hline
OQ4 & CEQ & In which phases of the software development lifecycle do you most frequently utilize bots? \\
\hline
OQ5 & CEQ & In your opinion, what are the primary benefits of using bots in software development? \\
\hline
OQ6 & CEQ & Have you ever encountered issues due to the use of bots in your projects? \\
\hline
OQ7 & OEQ & What enhancements would you like to see in the bots utilized in the software development process? \\
\hline
\end{tabular}
\label{tab3}
\end{center}
\end{table}

\subsection{Execution}
The questionnaire was distributed to professionals working in software engineering in the industry and to software engineering researchers. In order to reach participants from various locations across the country and the world, the invitation methods included professional chat channels on platforms such as WhatsApp, Telegram groups, post on X and direct outreach through LinkedIn. The survey was conducted from October 31, 2024, to November 6, 2024. Participation was voluntary, and the estimated time to complete the questionnaire was between 5 to 10 minutes. A total of 37 responses were obtained. The number of responses collected in a survey is an important factor regarding both the results and its validity. Other studies in the literature also had similar numbers of participants \cite{b25, b26}.

\subsection{Analysis}
The data obtained from the questionnaire were analyzed quantitatively. In the quantitative analysis, descriptive statistics was used to represent and describe the data to characterize the participants and to cross such information to obtain conclusions. 
After administering the questionnaire, the data were organized and analyzed both quantitatively and qualitatively. 
The quantitative analysis is discussed, as follows. In turn, a procedure inspired on Grounded Theory (GT) was used as the qualitative analysis procedure \cite{b26}. 
In this study, only two phases were utilized: (1) open coding and (2) axial coding. In the open coding, break, analysis, comparison, conceptualization, and categorization of the data are performed. A meticulous reading of the collected data is performed, and each text fragment receives an expression, sentence, or word, forming a code or category in the first steps of this phase. In the axial coding, relationships among categories and subcategories are established to form dense, related categories. The coding process is finished when no new data adds new knowledge to the categorization process. Despite the GT purpose for creating theories, Corbin \cite{b27}, explain that a researcher can use only a few steps to reach their research goals, e.g. when researchers only need to understand a specific phenomenon or situation.

\section{Results}

\subsection{Quantitative Analysis}
The first part of the questionnaire contains demographic questions to obtain the participants' profiles. Almost half of the respondents (18 out of 37 participants), representing 48.6\%, fall within the 25 to 34 age range. Participants from other age groups were also included: 7 participants (18.9\%) were aged 18 to 24, 9 participants (24.3\%) were aged 35 to 44, and 3 participants (8.1\%) were aged 45 to 54. There were no participants in the 16 to 18 or 55 and older age ranges. Another aspect identified was the country of residence. It was found that 33 participants (89.2\%) reside in Brazil, 2 participants reside in Spain (5.4\%), and Portugal and Switzerland each have 1 participant (2.7\%). Another aspect verified was the Brazilian states in which participants reside. It was found that 13 participants (35.1\%) reside in the state of Goiás, eight (21.6\%) in São Paulo, three (8.1\%) in Paraíba, and two (5.4\%) in Rio de Janeiro. Four participants (10.8\%) reside outside of Brazil, and each of the states of Amazonas, Espírito Santo, Minas Gerais, Piauí, Rio Grande do Sul, Santa Catarina, as well as the Federal District, has one participant (2.7\%) each. 

It was also analyzed how many years of experience the participants have in software development. The majority (14 out of 37 participants) have over 10 years of experience. Nine participants (24.3\%) have between 2 and 5 years, and another nine have 5 to 10 years of experience. Additionally, two participants (5.4\%) have 1 to 2 years of experience, and three (8.1\%) have less than 1 year. Another aspect verified was the current field of work. It was found that 28 participants (75.7\%) work directly in software development. Four participants (10.8\%) are engaged in DevOps/CI/CD, two (5.4\%) in data analysis or business intelligence, while the areas of quality assurance (QA), project management, and other (unspecified) each have one participant (2.7\%). The company size where each participant works was also verified. It was found that 12 participants (32.4\%) work in companies with over 1,000 employees, while companies with 501 to 1,000 employees and those with 1 to 10 employees each have 5 participants (13.5\%). Eight participants (21.6\%) work in companies with 51 to 200 employees, four (10.8\%) work in companies with 201 to 500 employees, and three (8.1\%) work in companies with 11 to 50 employees.

Next, the objective was to identify the frequency of bot usage in software development projects (OQ2). Participants were asked to classify their frequency of use as "Never," "Rarely," "Sometimes," "Frequently," or "Always". Figure 2 shows that only 3 participants (8.1\%) reported never using bots, while 14 participants (37.8\%) reported using bots frequently, indicating that the majority utilize bots in their projects.

\begin{figure}[htbp]
\centerline{\includegraphics[scale=0.6]{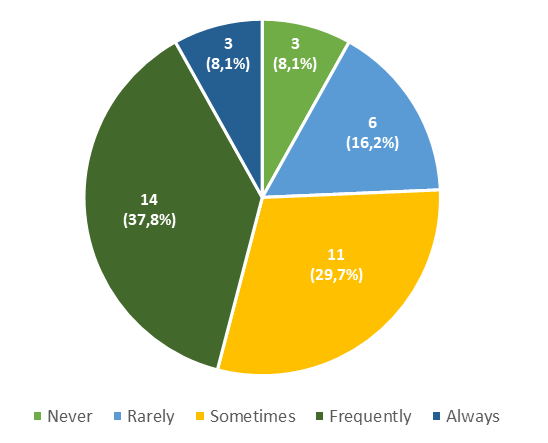}}
\caption{Frequency of Bot Usage in Software Development Projects}
\label{fig2}
\end{figure}

As the focus of the study is to identify software engineering professionals' perceptions regarding the use of bots in the software development process, a question using a Likert response scale (OQ3) ranging from "strongly disagree" to "strongly agree" was developed. As shown in Figure 3, respondents' answers ranged from a minimum of "disagree" to a maximum of "strongly agree." Additionally, the median response was identified as "agree," indicating a positive trend in the responses, with the majority of participants agreeing that bots increase productivity.

\begin{figure*}[htbp]
\centerline{\includegraphics[scale=0.4]{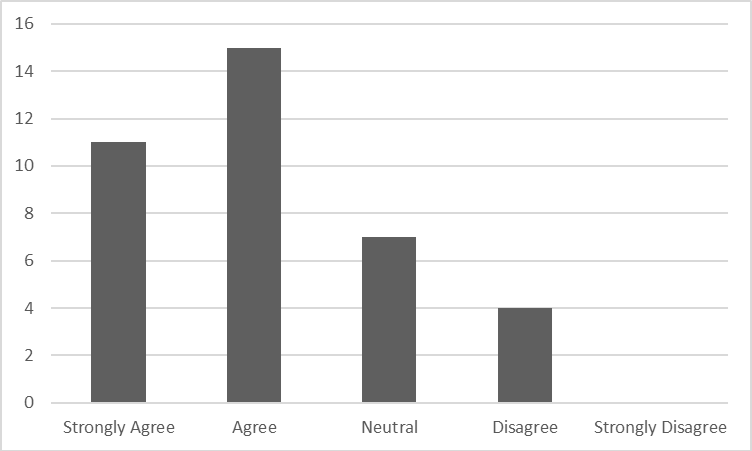}}
\caption{Likert Scale Responses to "Bots increase Productivity in Software Development}
\label{fig3}
\end{figure*}

\begin{figure*}[htbp]
\centering
\centerline{\includegraphics[scale=0.4]{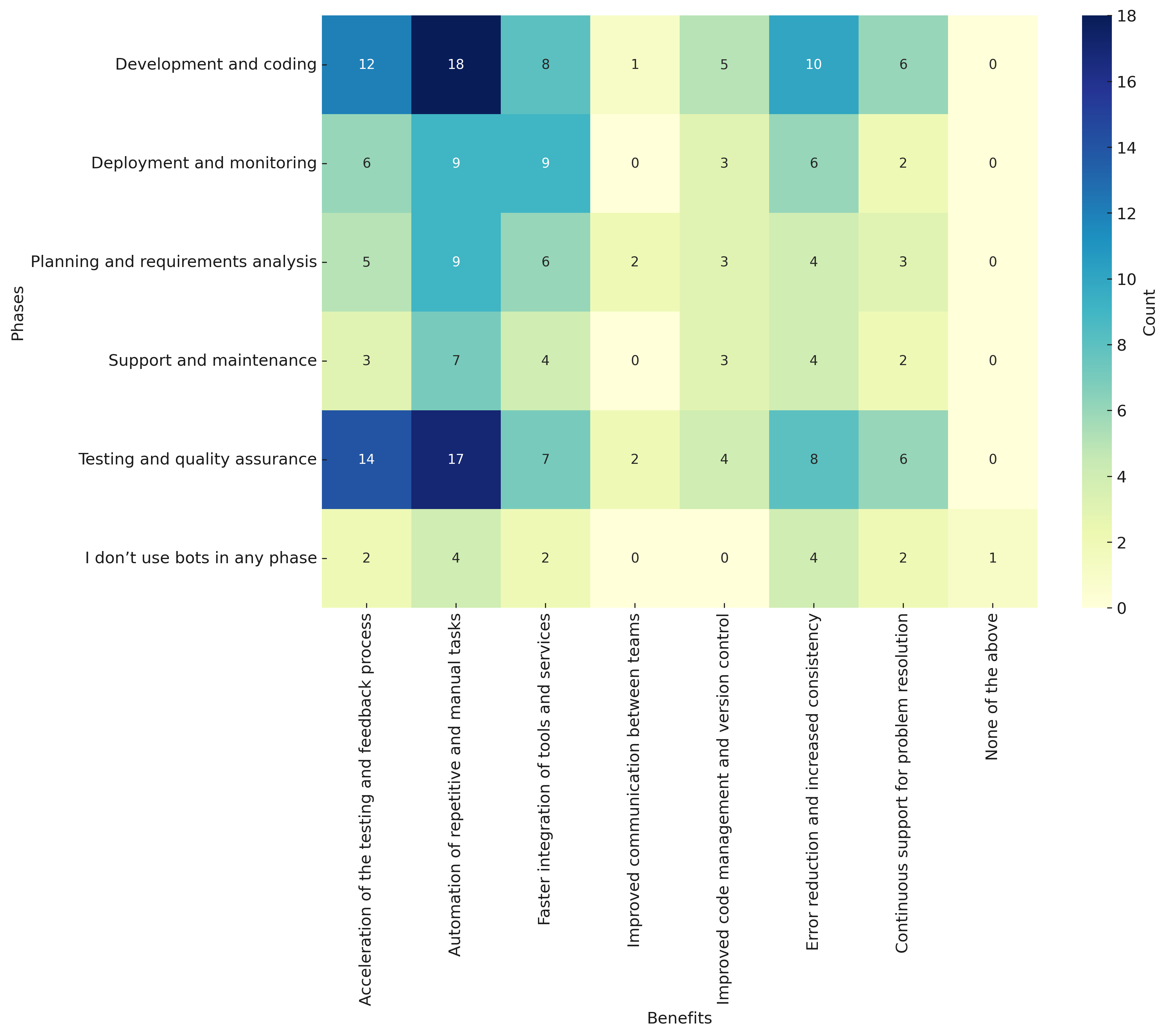}}
\caption{Heatmap of Software Development Phases Utilizing Bots vs. Usage Benefits}
\label{fig4}
\end{figure*}

Furthermore, it was also examined at which stage of the software development process bots can be most effectively utilized and the benefits that software engineers perceive from their use (OQ4 and OQ5). A data cross-reference was performed between the responses obtained in OQ4 and OQ5, as can be seen in the heat map in Figure 4. This figure illustrates the distribution of bot usage across various phases of the software development cycle, along with the associated benefits for each phase. It was found that the ``Development and Coding'' phase shows a high concentration of bot usage, especially for automating repetitive tasks (18 occurrences), indicating significant bot utilization to streamline coding workflows and maintain quality. Another phase with high concentration is ``Testing and Quality Assurance,'' where bot usage is essential for accelerating tests (14 occurrences), maintaining consistency, and automating repetitive tasks (17 occurrences). Conversely, the option "None of the alternatives" in OQ5 shows minimal or zero values in most rows, confirming that most respondents use bots for specific and tangible benefits, rather than finding no benefit at all.


Regarding bot usage, RQ6 indicated that 24 (64.9\%) participants had encountered no issues arising from the use of bots, showing that the majority of respondents experienced no problems due to bot usage.

\begin{figure*}[htbp]
\centerline{\includegraphics[scale=0.3]{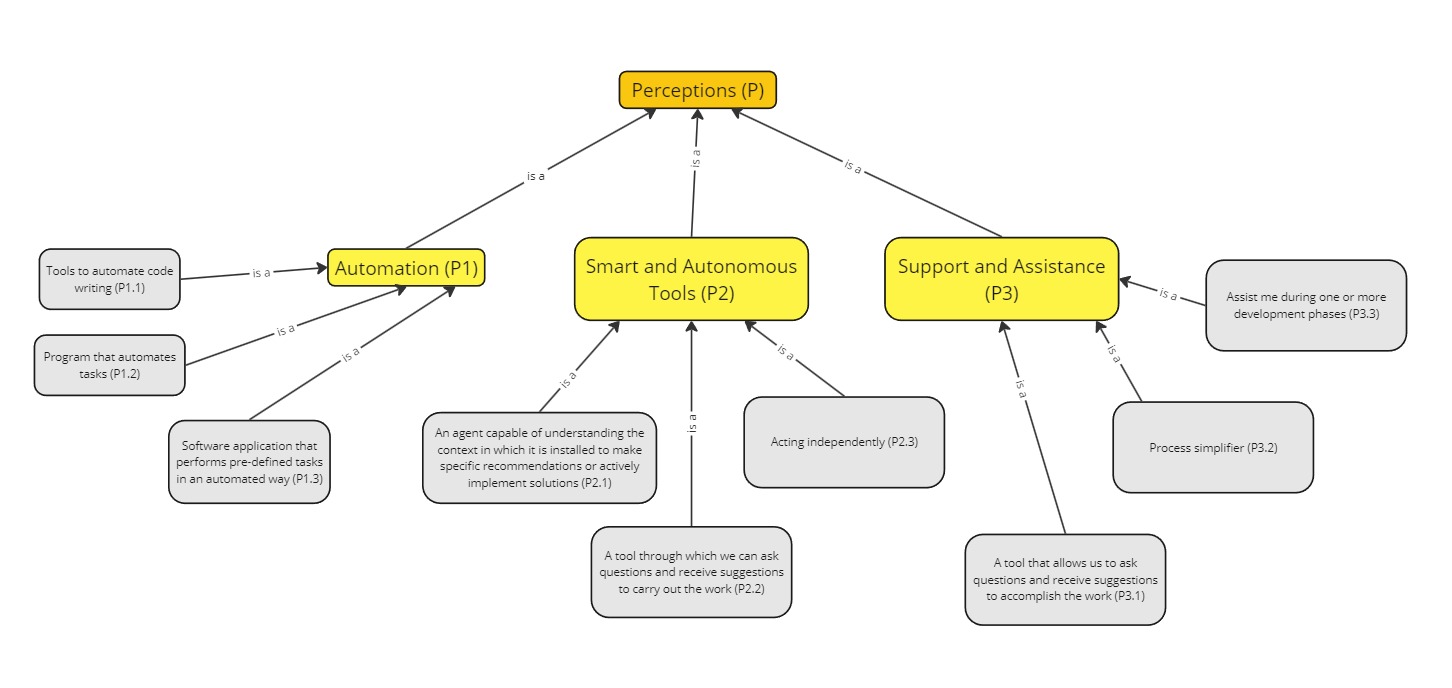}}
\caption{Some perceptions related to the definition of a bot}
\label{fig5}
\end{figure*}

\subsection{Qualitative Analysis}
The qualitative analysis was performed using questions OQ1 and OQ7. To do so, open coding and axial coding were applied, as mentioned in Subsection C of Section II. In open coding, the data were meticulously analyzed to create codes related to participants' responses. After the open coding phase, categories were identified, and relationships between codes were established, generating interrelationships. The open coding was conducted using the QualCoder\footnote{https://qualcoder.wordpress.com/} tool, while the graphical representation of the coding was created using Miro\footnote{https://miro.com/pt/}.

In the qualitative analysis of OQ1, categories related to perceptions of what constitutes a bot were identified: (i) Automation, (ii) Smart and Autonomous Tools, and (iii) Support and Assistance. Figure 5 provides a detailed breakdown of perceptions related to these categories. It was observed that bots are perceived as automated tools that perform and facilitate specific, often repetitive tasks that would otherwise require human time and effort.

\begin{figure*}[htbp]
\centerline{\includegraphics[scale=0.3]{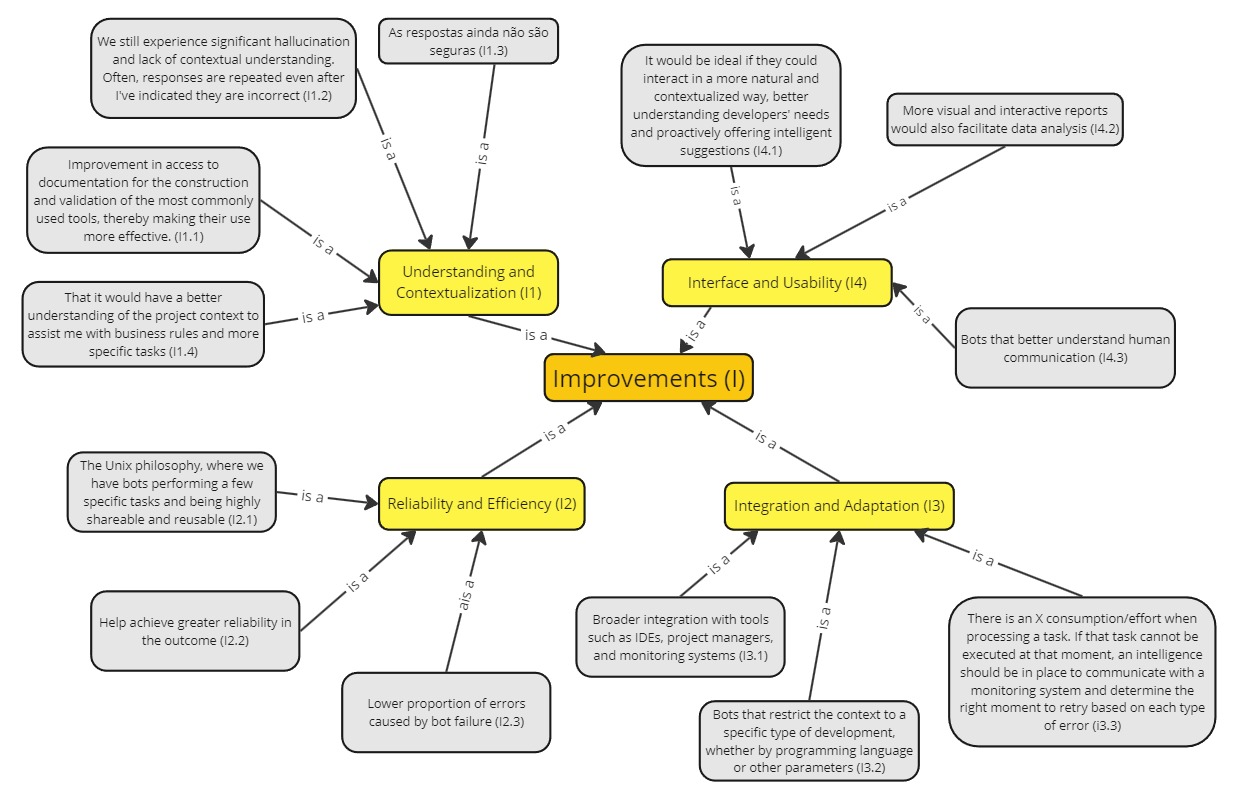}}
\caption{Some suggestions for improvements related to bots}
\label{fig6}
\end{figure*}

In the qualitative analysis of RQ7, four categories were identified regarding suggested improvements for bots: (i) understanding and contextualization, (ii) reliability and efficiency, (iii) integration and adaptation, and (iv) interface and usability. Figure 6 presents some of the improvements suggested by software engineers within these categories. The analysis revealed that participants recognize the potential of bots to increase efficiency; however, they desire enhancements that improve bots' reliability, situational context, and integration with other tools, as well as a more streamlined and intuitive user experience. These improvements would make bots more adaptable tools, aligned with the specific demands of development.

The analysis of the results obtained from the survey of software engineering professionals, along with the subsequent evaluation, provided insights into professionals' perceptions of bot usage in the software development process and allowed for answering the two research questions, RQ1 and RQ2. Regarding RQ1, it was identified that software engineering professionals perceive bots as essential tools for automation and assistance in software development. They are seen as tools that not only accelerate software development but also offer extensive support to professionals, especially when combined with artificial intelligence (AI). On the other hand, for RQ2, it was found that, although bots have a significant impact on the routine of software engineering professionals, they still require numerous improvements to further adapt to their users or collaborators, particularly in terms of understanding and contextualization, reliability and efficiency, integration and adaptability, as well as interface and usability.

\noindent\textbf{Threats to Validity.} Although a significant (albeit limited) sample of 37 software engineering professionals was obtained, we sought to maximize the number of participants by inviting as many national and international professionals as possible within a fixed timeframe. Nevertheless, the study was undoubtedly constrained by the number of responses. It is also important to highlight that the questionnaire structure, being developed by the researchers, may inherently contain some bias.

\section{Conclusion}
The main contribution of this study was to report findings addressing the research question \textit{What are the perceptions of software engineers regarding the use of bots in the software development process?}". Survey responses were analyzed and results were synthesized. Out of a total of 37 participants, nearly half agreed that bots enhance productivity in the software development process. 
In conclusion, this study underscores the significance of using bots in the software development process to increase software engineers' productivity.


\begin{thebibliography}{00}
\bibitem{b1} Eric Raymond. The cathedral and the bazaar. Knowledge, Technology \& Policy, 12(3):23–49, 1999.
\bibitem{b2} E. Raymond. The catherdral and the bazaar. First Monday, http://www.firstmonday.dk/, v. 3, 1998.
\bibitem{b3} J. Feller and B. Fitzgerald. Understanding Open Source Software Development. Addison-Wesley, London, UK, 2002.
\bibitem{b5} Mehdi Golzadeh, Damien Legay, Alexandre Decan, and Tom Mens. 2020. Bot or not? Detecting bots in GitHub pull request activity based on comment similarity. In Proceedings of the IEEE/ACM 42nd ICSEW'20. ACM, New York, NY, USA, 31–35. 
\bibitem{b9} Carlene Lebeuf, Margaret Anne Storey, and Alexey Zagalsky. 2017. Software Bots. IEEE Software 35, 1 (2017), 18–23. 
\bibitem{b11} Mairieli Wessel, Bruno Mendes De Souza, Igor Steinmacher, Igor S. Wiese, Ivanil-ton Polato, Ana Paula Chaves, and Marco A. Gerosa. 2018. The power of bots: Understanding bots in OSS projects. Proc. ACM on Hum-Comput. Interact. (2018). 
\bibitem{b14} Wessel, Mairieli \& Serebrenik, Alexander \& Wiese, Igor \& Steinmacher, Igor \& Gerosa, Marco Aurelio. (2020). Effects of Adopting Code Review Bots on Pull Requests to OSS Projects. 2020 ICSME.
\bibitem{b15} M. Golzadeh, A. Decan, E. Constantinou and T. Mens, "Identifying bot activity in GitHub pull request and issue comments," in 2021 IEEE/ACM Third BotSE, Madrid, Spain, 2021 pp. 21-25. 
\bibitem{b16} Mairieli Wessel, Igor Steinmacher, Igor Wiese, and Marco A. Gerosa. 2019. Should I stale or should I close? an analysis of a bot that closes abandoned issues and pull requests. In Proceedings of the 1st BotSE '19. IEEE Press, 38–42. 
\bibitem{b18} Margaret-Anne Storey and Alexey Zagalsky. 2016. Disrupting developer productivity one bot at a time. In Proceedings of the 2016 24th ACM SIGSOFT-FSE 2016. Association for Computing Machinery, New York, NY, USA, 928–931.
\bibitem{b19} Araújo Lebtag, B. G., Teixeira, P. G., dos Santos, R. P., Viana, D., \& V. Graciano Neto, V. (2022). Strategies to Evolve ExM Notations Extracted from a Survey with Software Engineering Professionals Perspective. JSERD, 10, 2:1 – 2:24.
\bibitem{b20} Wohlin, C., Runeson, P., Hst, M., Ohlsson, M. C., Regnell, B., and Wessln, A. (2012).Experimentation in Software Engineering. Springer Publishing Company, Incorporated.
\bibitem{b21} Molléri, J. S., Petersen, K., and Mendes, E. (2016). Survey guidelines in software engineering: An annotated review. In Proc. of ESEM ’16, pages 58:1–58:6.
\bibitem{b22} Lebtag, B., Teixeira, P., Santos, R., Viana, D., and Graciano Neto, V. (2020). Evaluating the understandability and expressiveness of simulation executable models with professionals – obtaining perceptions from researchers and practitioners for improving quality of models. In 19th SBQS’20, New York, NY, USA. ACM.
\bibitem{b23} Simon Urli, Zhongxing Yu, Lionel Seinturier, and Martin Monperrus. 2018. How to design a program repair bot? insights from the repairnator project. In Proceedings of the 40th ICSE-SEIP '18. ACM, New York, NY, USA, 95–104. 
\bibitem{b24} I. Beschastnikh, M. F. Lungu and Y. Zhuang, "Accelerating Software Engineering Research Adoption with Analysis Bots," 2017 IEEE/ACM 39th ICSE-NIER, Buenos Aires, Argentina, 2017, pp. 35-38.
\bibitem{b25} A. Pinheiro, C. Rabello, L. Furtado, G. Pinto, and C. Souza. "How Do Bot Developers Perceive Bot Development? A Survey", in SBC, Belém, 2019, pp. 11-20.
\bibitem{b26} Mairieli Wessel, Alexander Serebrenik, Igor Wiese, Igor Steinmacher, and Marco A. Gerosa. 2020. What to Expect from Code Review Bots on GitHub? A Survey with OSS Maintainers. In Proc. of XXXIV SBES '20. ACM, New York, NY, USA, 457–462. 
\bibitem{b27} Corbin,J.(2015).Basics of qualitative research : techniques and procedures for developing grounded theory. SAGE.
\end{thebibliography}
\end{document}